\documentclass[twocolumn]{aastex631}
\usepackage{booktabs} 
\usepackage{tablefootnote} 
\usepackage{threeparttable}
\usepackage{amsmath}

\defcitealias{Fox10}{F10}
\defcitealias{Sembach03}{S03}

\begin{document}

\title{The Observed \ion{O}{6} is Just the Tip of the Iceberg: Estimating the Hidden Material in Circumgalactic and Intergalactic Clouds}

\author[0009-0007-3193-0090]{Eric Goetz}
\email{eric.goetz@uga.edu, chen.wang3@uga.edu, robinl1@uga.edu}
\affiliation{Department of Physics and Astronomy and Center for Simulational Physics, University of Georgia, Athens, GA, 30602, USA}

\author[0009-0009-2215-3941]{Chen Wang}
\affiliation{Department of Physics and Astronomy and Center for Simulational Physics, University of Georgia, Athens, GA, 30602, USA}

\author[0000-0001-5221-0315]{Robin Shelton}
\affiliation{Department of Physics and Astronomy and Center for Simulational Physics, University of Georgia, Athens, GA, 30602, USA}



\begin{abstract}
This paper proposes a new method for estimating the total quantity of material in moving circumgalactic and intergalactic clouds from \ion{O}{6} measurements.  
We simulate high-velocity clouds (HVCs) with the FLASH hydrodynamic code and track the ionization and recombination of all ionization levels of oxygen as a function of time.  
We calculate the \ion{O}{6}/oxygen ratio ($f_{\rm O VI}$) in our dynamic NEI clouds, finding that it differs significantly from that in static gas.
We find that \ion{O}{6} exists in cool, medium, and hot gas in the   clouds.  
As such, it traces all of the hydrogen rather than merely the ionized hydrogen.
The total quantity of hydrogen along a typical observed line of sight through a cloud can be estimated from the observed \ion{O}{6} column density, metallicity, and our $f_{\rm O VI}$.
We provide the simulations' $f_{\rm O VI}$, a prescription for finding $f_{\rm O VI}$ for observed dynamic clouds, and a methodology for calculating the total hydrogen column density from this $f_{\rm O VI}$ and an observed \ion{O}{6} column density.
As examples, we use our $f_{\rm O VI}$ to estimate the total hydrogen column densities along various observed sight lines through two HVCs, Complex C and the Magellanic Stream, finding that these clouds contain more material than the previous lower limits. 
 We also extend this analysis to {low-redshift} intergalactic {\ion{O}{6}} clouds, finding that they contain several times more baryonic material than previously thought and therefore may account for a significant fraction of the Universe's baryons.  
 \end{abstract}



\section{Introduction} \label{sec:intro}
Gaseous clouds are ubiquitous across the Universe. They are found in nearly every type of structure, from protostellar objects to galaxy clusters.  They are of obvious importance to their hosts.  Clouds in and near our Galaxy, for example, affect the star formation rate and galactic evolution while clouds throughout intergalactic space contribute to the tally of baryonic mass in the Universe \citep{Richter04}.  Examples of the former are   {high-velocity clouds} (HVCs), clouds with \mbox{$|v_{LSR}| \geq 90$ {km~s$^{-1}$}} \citep{Wakker97} or \mbox{100 km~s$^{-1}$} \citep{Richter17}.  They affect our Galaxy’s evolution by providing gas that can be used in the future to form additional stars \citep{Putman12, Fox19}.

In general, HVCs are multi-structured and multiphase. They can contain neutral gas such as \ion{H}{1} and \ion{O}{1}, but can also contain low ions such as \ion{H}{2}, \ion{C}{2} and \ion{O}{2}, and even high ions such as \ion{C}{4}, \ion{O}{6} and \ion{Si}{4}. 

Many HVCs appear to be out of collisional ionization equilibrium (CIE).  This is because their absorption features have such narrow line widths that their temperatures must be well below their CIE temperatures (e.g.,\citealt{Yao11}; \citealt{Tripp22}).  Photoionization has been suggested as a possible way to highly ionize the gas without substantially heating it. However, the photon field at 114 eV is generally considered too weak to ionize substantial amounts of \ion{O}{5} to \ion{O}{6} (\citealt{Tripp03}, \citealt{Sembach03}, hereafter \citetalias{Sembach03}, \citealt{Fox04}, \citealt{Ganguly05}, \citealt{Fox10}, hereafter \citetalias{Fox10}).
Cooling flow models have also been considered \citep{Heckman02, Tripp22}, because they naturally explain high ions at temperatures below their CIE temperatures. The mixing between hot ambient gas and cooler HVC gas will also yield large numbers of cool high ions \citep{Kwak10}. 

Additional indications that the gas is out of CIE come from line ratios, such as \ion{Si}{4}/\ion{C}{4} and \ion{C}{4}/\ion{O}{6}. 
 Simultaneous fitting of these line ratios shows that the predictions of non-equilibrium ionization (NEI) models are closer to observational data than those from CIE models \citep{Tripp22}.

Of the high ions present in HVCs, \ion{O}{6} is often chosen as a tracer of highly ionized HVC gas (\citetalias{Sembach03}; \citealt{Fox06};  \citetalias{Fox10}; \citealt{Cashman23}).
because oxygen is the most abundant metal in HVCs and because its large oscillator strength allows for easier observations \citep{Morton91}. The Hubble Space Telescope and the Far Ultraviolet Spectroscopic Explorer have detected \ion{O}{6} in HVCs along sight lines toward many AGNs. 

Once an \ion{O}{6} column density has been observed, it can be used to estimate the associated hydrogen column density if the oxygen metallicity and \ion{O}{6} ionization fraction (\ion{O}{6}/oxygen, $f_{\rm O VI}$) of the cloud are also given. In the past, the hydrogen associated with \ion{O}{6} had been assumed to be hot and ionized (e.g. \citetalias{Sembach03}; \citealt{Sembach04}; \citetalias{Fox10})  because according to CIE and NEI calculations of static gas \citep{Sutherland93, GS07}, \ion{O}{6} only exists at temperatures {on the order of} $10^5$~K. 
Once calculated, the resulting \ion{H}{2} column density was then added to the \ion{H}{1} column density measured from observations of the 21 cm lines or the hydrogen Lyman series lines, if available, to estimate the total hydrogen column density.
Some authors also added components related to \ion{Si}{4} and/or other ions (e.g. \citetalias{Fox10}). 
Until now, calculations of the \ion{H}{2} column density have adopted as an upper limit the maximum value of $f_{\rm OVI}$ from static gas models such as \citet{Sutherland93,GS07}.
The peak $f_{\rm OVI}$ in those models is $\sim0.2$ and occurs at a temperature of $\sim3 \times 10^5$ K.

We set about to calculate the average $f_{\rm OVI}$ for dynamic NEI gas, with the awareness that since some \ion{O}{6} exists at temperatures above and below $3 \times 10^5$ K, the average $f_{\rm OVI}$ will be below the peak value.
This new value of $f_{\rm O VI}$ results from FLASH simulations of the ionization and recombination of oxygen in HVCs. The FLASH code \citep{Fryxell00} has been used previously to study HVCs \citep{Kwak11,Ploeckinger12,Gritton14,Gritton17,Galyardt16,Sander21}. 
Considering the variety of individual HVCs and the uncertainty in the density, temperature, size, and velocity of specific HVCs, we simulate ten HVCs spanning a range of initial cloud parameters. We track their evolutions over time as they interact with the ambient medium.  The dynamic nature of these simulations allows for mixing between the cold cloud and the hot ambient material.  This mixing allows ambient material to become entrained within the cloud.  These interactions, combined with the non-equilibrium nature of ionization and recombination, have significant effects on the ionization fraction of \ion{O}{6}.

We find that \ion{O}{6} physically overlaps with both neutral and ionized hydrogen in the cloud.  Therefore, the entire cloud must be considered when finding the average $f_{\rm OVI}$.
We calculate $f_{\rm O VI}$ from the ratio of the number of \ion{O}{6} ions in the cloud to the number of {oxygen atoms} in the cloud.
The hydrogen column density calculated from this $f_{\rm OVI}$ and \ion{O}{6} measurements must therefore be reinterpreted.
It is the hydrogen in all phases rather than merely the \mbox{$\sim3 \times 10^5$ K} component of the hydrogen.

In our simulations, the ratio of neutral hydrogen to \ion{O}{6} (i.e., \ion{H}{1}/\ion{O}{6}) decreases over time as the cold cloud material mixes with the hot ambient gas. The \ion{H}{1}/\ion{O}{6} ratio can vary from cloud to cloud, or even across a single cloud. 
The $f_{\rm OVI}$ ratio evolves as the cloud evolves, but it is correlated to the \ion{H}{1}/\ion{O}{6} ratio. 
We present the relationship between \ion{H}{1}/\ion{O}{6} and $f_{\rm OVI}$ in Figure \ref{fig:pres} in the Discussion (Section \ref{sec:discussion}). From this figure, the appropriate value of $f_{\rm OVI}$ can be determined for any given sight line from the observed \ion{H}{1}/\ion{O}{6} ratio.

This paper develops a methodology for calculating the total amount of hydrogen from an observed \ion{O}{6} column density. 
This new methodology requires an estimate of the average $f_{\rm OVI}$ across the cloud, which we have extracted from the simulations presented in this paper.
We present a prescription for determining the $f_{\rm OVI}$ pertaining to a given line of sight from the observed column densities of \ion{O}{6} and \ion{H}{1} on the sight line.
 
 We describe our setup and state our simulation parameters for our 10 FLASH simulations in Section \ref{sec:methods}.  In Section \ref{sec:results}, we present the results of the simulations and the post-processing steps we take.  
 We also present our updated methodology for calculating the total hydrogen column density and apply it to two example clouds: Complex C and the Magellanic Stream (MS).  Our method yields more material than do previous methods based on the peak value of the static CIE or NEI $f_{\rm OVI}$.  
 In Section \ref{sec:discussion} we present our prescription for determining $f_{\rm OVI}$ from observed \ion{H}{1}/\ion{O}{6} ratios on other sight lines and show that there is much more material in HVCs and {low-redshift} intergalactic {\ion{O}{6}} clouds than previously thought.

\section{Methods} \label{sec:methods}
We run our simulations with FLASH version 4.6.2 \citep{Fryxell00}. We use FLASH’s hydrodynamic and NEI modules to calculate the time-dependent ionization levels of oxygen and we use a lookup table to calculate the radiative cooling of the materials. For the latter, we use the CIE cooling curve table with \mbox{[Fe/H]=-0.5} from \citet{Sutherland93}. 

The simulations are done in 3D cartesian coordinates with adaptive mesh refinement. In each simulation, the domain is a rectangular cuboid box with dimensions of 2.4 kpc $\times$ 1.2 kpc $\times$ 10.8 kpc in the $\hat{x}$, $\hat{y}$, and $\hat{z}$ directions, respectively. The domain is segmented into 18 identical square blocks in a 2 $\times$ 1 $\times$ 9 layout. The maximum resolution of each cell is 9.375 pc in each direction.

We run the simulations in a wind tunnel fashion, with the ambient material flowing in through the lower z boundary and out through the upper z boundary. Material can also leave the domain through the lower and upper x boundaries and upper y boundary. In order to capitalize on the assumed symmetry of the cloud, we place the center of the cloud along the lower y boundary, simulate half of the cloud inside the domain, and make the lower y boundary a reflecting boundary. The center of the cloud is initially located 1.2 kpc above the lower z boundary. This location is at (x=0 kpc, y=0 kpc, z=0 kpc) in the domain.  From the perspective of the domain, the cloud is stationary at the beginning of the simulation. However, from the perspective of the ambient gas, the cloud is moving toward the $-\hat{z}$ direction with a velocity equaling the wind speed.

In this project, we use \ion{O}{6} as a tracer of hydrogen in highly ionized gas, so it is important to use a reasonable estimate of the oxygen metallicity in our calculations.
We want to choose a metallicity that is similar to those of our comparison clouds, the MS and Complex C.
The oxygen metallicity varies throughout the MS and Complex C (\citealt{Fox13B}; \citetalias{Sembach03}) and different observers find somewhat different values. 
A metallicity value of 0.1 solar was adopted for calculations by both \citetalias{Fox10} for the MS and \citetalias{Sembach03} for Complex C.  \citetalias{Fox10} used the \citet{Asplund09} abundance table in which there are $4.89 \times 10^{-4}$ oxygen atoms per hydrogen atom.  \citetalias{Sembach03} used the \citet{Holweger01} abundance table in which there are $5.45 \times 10^{-4}$ oxygen atoms per hydrogen atom, but they calculate the total hydrogen mass of the cloud instead of hydrogen column densities along sight lines. To be consistent with the calculations of \citetalias{Fox10}, we adopt the same metallicity value and abundance table that they used.
This resulting value of O/H is similar to those of \citet{Fox13A} and \citet{Howk17} for the MS and \citet{Collins07} for Complex C. Although \citet{Wakker99} and \citet{Richter01} also found a 0.1 solar metallicity, they used the \cite{Anders89} abundance table in which there are $\sim1.6$ times more oxygen atoms per hydrogen atom.  
Other observations of Complex C suggest a higher metallicity, up to 0.3 solar \citep{Tripp03, Shull11}, using \cite{Holweger01} and \cite{Asplund09}, respectively. Considering this, we also examine the effect of using alternate cloud metallicities.

A wide range of values for the metallicity of the ambient medium have been deduced.
\citet{Miller15} used \ion{O}{8} emission lines to calculate a metallicity of the circumgalactic medium that was $\geq 0.3$ solar assuming \cite{Anders89} abundances.  
\citet{Troitsky17} modeled the density of the circumgalactic gas, finding the best fit metallicity at a height of 10 kpc to be roughly 0.5 solar, with significant uncertainties. They appear to have relied on \cite{Miller15}, which uses \cite{Anders89} abundances.   
\citet{Miller16} examined \ion{O}{7} absorption lines and pulsar dispersion measures and found the value of the halo gas metallicity to be $\geq 0.6$ solar. They used \cite{Holweger01} abundances.
Comparisons of observations of the halo with simulations suggest that the metallicity of the halo is around 0.7 solar \citep{Henley17}. They used FLASH, which usually uses \cite{Anders89} abundances.   
\citet{Henley15} modeled the halo's x-ray emission using solar metallicity with a variety of oxygen abundance tables and found good reduced $
\chi^2$ values for all of them.  This suggests that the oxygen abundance of the halo is difficult to constrain.
Considering the range of deduced metallicities, we feel comfortable using a 0.7 solar metallicity.  However, we also explore the effect of using alternate ambient metallicities.

To implement these metallicities, we conduct a two-step process. The first step is to arbitrarily set the metallicities of the cloud and ambient gas to simple values in the FLASH simulations.  For convenience, we set the initial values for the cloud and ambient metallicities to $10^{-3}$ and 1 times the solar metallicity, respectively.  The second step is to re-scale them during the post-processing to 0.1 solar for the cloud and 0.7 solar for the ambient material. 
As mentioned above, we also re-scale to other metallicities to explore the impact on our results.
FLASH uses the abundance table of \cite{Anders89}, but we re-scale during the post-processing to the abundance table of \cite{Asplund09}.

\begin{table*}[t]
    \centering
    \caption{Simulation   {Parameters}}
    \begin{tabular}{lcccccc}
      Simulation & $n(\text{H})_{\text{cloud}}$ & $T_{\text{cloud}}$ & $n(\text{H})_{\text{ambient}}$ & $T_{\text{ambient}}$ & $r_{\text{cloud}}$ & $v_{\text{inflow}}$\\ 
                 & (cm$^{-3}$) & (K) & (cm$^{-3}$) & (K) & (pc) & (  {km~s$^{-1}$})\\
      \hline           
      Run 1 & 0.4 & 5000 & 0.001 & $2\times10^6$ & 500 & 150 \\ 
      Run 2 & 0.4 & 5000 & 0.001 & $2\times10^6$ & 500 & 100 \\ 
      Run 3 & 0.4 & 5000 & 0.001 & $2\times10^6$ & 300 & 150 \\ 
      Run 4 & 0.4 & 5000 & 0.001 & $2\times10^6$ & 500 & 300 \\ 
      Run 5 & 0.04 & 5000 & 0.0001 & $2\times10^6$ & 500 & 150\\ 
      Run 6 & 0.2 & 5000 & 0.001 & $1\times10^6$ & 500 & 150 \\ 
      Run 7 & 2.0 & 1000 & 0.001 & $2\times10^6$ & 500 & 150 \\ 
      Run 8 & 1.0 & 1000& 0.001 & $1\times10^6$ & 500 & 150 \\ 
      Run 9 & 0.67 & 3000 & 0.001 & $2\times10^6$ & 500 & 300 \\ 
      Run 10 & 0.222 & 9000 & 0.001 & $2\times10^6$ & 500 & 150 \\ 
    \end{tabular}
    \label{tab:params}
\end{table*} 

In order to explore how the cloud and ambient materials mix, cool, ionize, and recombine, we run 10 simulations with a variety of initial densities, temperatures, radii, and velocities (see Table \ref{tab:params}.) 
Among our set of simulations, our choice of cloud hydrogen density, $n(\text{H})_{\text{cloud}}$, ranges from 0.04 cm$^{-3}$ to 2 cm$^{-3}$. Our choice of cloud temperature, $T_{\text{cloud}}$, ranges from 1000 K to 9000 K. The ambient temperature, $T_{\text{ambient}}$, is set to $1\times10^6$ K or $2\times10^6$ K. The ambient density, $n(\text{H})_{\text{ambient}}$, is set to $1\times 10^{-3}$ cm$^{-3}$ or $1\times 10^{-4}$ cm$^{-3}$. 
We start   {each} simulation with the cloud in pressure equilibrium with the ambient material. 
This constrains the cloud density,   {considering} the cloud temperature, ambient temperature, and ambient density   {have already been} specified. Our preliminary simulations and other work \citep{Gronnow17, Gritton17} have shown that the initial radius of the cloud affects the cloud’s evolution, mixing and ionization levels. 
For this reason, we run simulations with two choices of cloud radius, 300 pc and 500 pc. In order to observe how the inflow of ambient gas affects its mixing with the cloud material, we also   {adopt} three different wind speeds, \mbox{100 km~s$^{-1}$}, \mbox{150 km~s$^{-1}$}, and \mbox{300 km~s$^{-1}$}. The simulations are run for 200 Myr, so each cloud has enough time to mix with the ambient gas and evolve into a more realistic shape.
However, due to computational limitations, Runs 4 and 5 stopped after 185 and 128 Myr, respectively.  Both simulations ran long enough for their clouds to fully evolve.

The cloud parameters given in Table \ref{tab:params} are the values at the center of the cloud. Near the outer region of the cloud, the density and temperature grade into those of the ambient material. The density follows the hyperbolic tangent function given by \citet{Gritton14}, but using a scale length of 50 pc. 
The temperature rises inversely with the density to preserve pressure equilibrium and mesh with the surrounding conditions. The transition between the cloud abundance and the ambient abundance occurs where the density is equal to $0.99 n_{\text{ambient}}+0.01 n_{\text{cloud}}$. The velocity transition occurs at the same place.

At the start of the simulation, the ambient material travels at the chosen wind speed, but the cloud is at rest.  As time progresses, some material is torn from the cloud while some ambient material is entrained into the cloud.  In general, mixing between cloud and ambient material causes some computational cells to contain similar proportions of both cloud and ambient gas.  This mixing erases the clear boundary between cloud and ambient material.  Therefore, a new criterion for defining cloud material must be found, one that is both consistent with observational analyses and computationally feasible.  
In observations of actual clouds, the velocity contrast between the cloud and ambient gas is used to identify the cloud material.  
Actual HVCs are defined as clouds whose velocity differs from the local standard of rest by at least 90 or 100  km~s$^{-1}$ \citep{Wakker97, Richter17}.  We use these ideas to develop a velocity criterion for identifying the cloud.  Any material in our domain whose velocity in the $\hat{z}$ direction differs by more than \mbox{100 km~s$^{-1}$} from that of the ambient material is considered to be part of the HVC.  The ambient material speeds up slightly as the simulation progresses, which is taken into consideration in the application of the velocity criterion.

Our simulations model the mass of hydrogen and helium in the gas. Other elements make inconsequential contributions to the mass. Each simulation tracks the ionization level populations of helium and oxygen in every cell in a time-dependent manner based on the collisional ionization and recombination rates. It does not track the ionization level populations of hydrogen. Therefore, during the post-processing, we assume that the ratio of neutral hydrogen atoms to all hydrogen atoms and ions is equal to the ratio of neutral oxygen atoms to all oxygen atoms and ions. We make this assumption because their first ionization potentials are almost identical \citep{Allen} and because there is rampant charge exchange \citep{Field71}.  

In reality, HVCs are bathed in a photon field that   {contributes to the photoionization} of each cloud. In order to estimate the extent of the photoionization, we perform Cloudy \citep{Ferland17} simulations during the post-processing. The photon field in the Cloudy simulations is the sum of the extragalactic background (EGB) and the radiation escaping from the Milky Way (MW). We use the HM05 \citep{Ferland17} table for the EGB contribution and use the \citet{Fox05} MW model for the MW contribution. 
The total ionizing flux for each contribution is taken from   \citetalias{Fox10}.
The abundance table used in our calculation \citep{Asplund09} is the same as that used in \citetalias{Fox10}.
Our technique was verified by reproducing parts of Figure 5 in   \citetalias{Fox10}, which used the same photon field model. In Cloudy, we model each cloud as a slab that has the same initial density and temperature as the cloud in the corresponding FLASH simulation.
We calculate the column density of hydrogen which is photoionized by the EGB and MW photon fields. We use this column density of hydrogen in our post-processing calculations in Section \ref{sec:results}.

\section{Results} \label{sec:results}

\subsection{Simulation Results}

\begin{figure}[ht]
    \centering
    \includegraphics[scale=0.5]{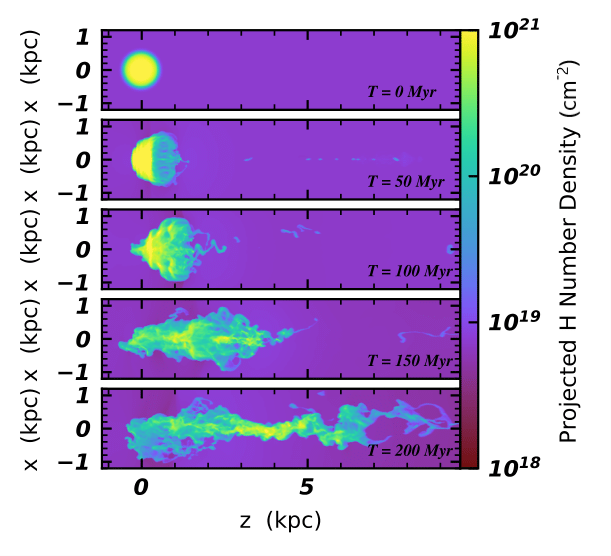}
    \caption{Hydrogen column density plots for Run 1 at 5 epochs (0, 50, 100, 150, and 200 Myr).  The column densities are calculated by integrating the hydrogen number density along sight lines directed into the page.  These sight lines are in the $\hat{y}$ direction, which is the direction in which only half the cloud was simulated.  So, we have multiplied the column densities by 2 to account for having simulated only 1/2 of the cloud.  The ambient gas moves to the right, interacting with the cloud gas.  The resulting tail formation and hydrodynamic instabilities are apparent in this figure.}
    \label{fig:hcd}
\end{figure}

\begin{figure}[ht]
    \centering
    \includegraphics[scale=0.5]{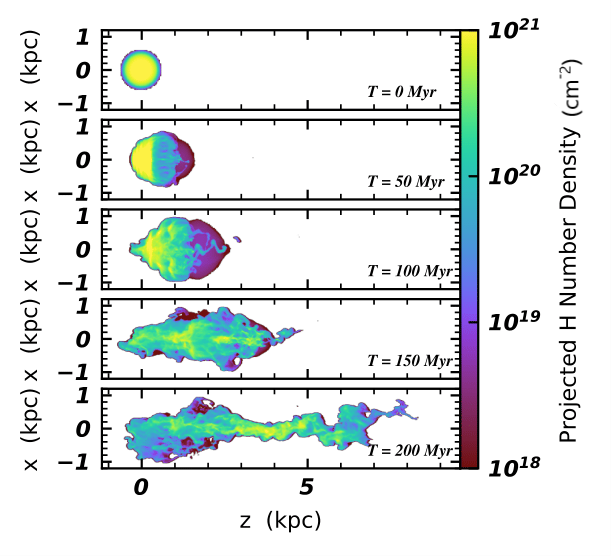}
    \caption{Hydrogen column density plots for Run 1 at 5 epochs (0, 50, 100, 150, and 200 Myr) for material whose velocity in the $\hat{z}$ direction differs by more than \mbox{100 km~s$^{-1}$} from that of the ambient material.  As in Figure \ref{fig:hcd}, the column densities include a factor of 2.  At late times, part of the cloud’s tail has slowed too much to be considered part of the cloud.  This can be seen by comparing Figure \ref{fig:hcd_cut} with Figure \ref{fig:hcd}. }
    \label{fig:hcd_cut}
\end{figure}

First, we will describe the clouds’ dynamics. As the simulations progress, material is ablated from the cloud.  This deforms the cloud from its initial spherical shape into a more elongated structure.  The ablated material forms a tail, as can be seen in Figure \ref{fig:hcd} for a representative simulation (  {Run 1} in Table \ref{tab:params}).  All figures and analyses were done with the YT software package \citep{YT}. The interaction between the ambient gas and the tail decelerates the tail.   Over time, the tail will approach the speed of the ambient gas   {and so will no longer meet the cloud criterion.}  This is shown by comparing Figure \ref{fig:hcd} with Figure \ref{fig:hcd_cut} for which the velocity criterion has been applied to the domain.  

Hydrodynamic instabilities such as the Rayleigh-Taylor and Kelvin-Helmholtz instabilities also affect the cloud.  Rayleigh-Taylor instabilities are due to the head-on collisions between cloud and ambient material \citep{Drazin02}.  Kelvin-Helmholtz instabilities are due to the velocity shear at the boundary of the cloud and ambient material \citep{Drazin02}.  These instabilities tear material from the cloud and incorporate hot, ambient material into the cloud.  Because the ambient material is very hot, it contains high ions of oxygen such as \ion{O}{7} to \ion{O}{9}.  Once entrained in the cloud, the hot gas transfers heat to the cold gas and both components radiatively cool.  Consequently, the high ions from the entrained ambient gas begin to recombine and the neutral and low ions in the cloud begin to ionize \citep{Kwak10, Kwak11}.

\begin{figure*}[ht]
    \centering
\includegraphics[scale=0.3]{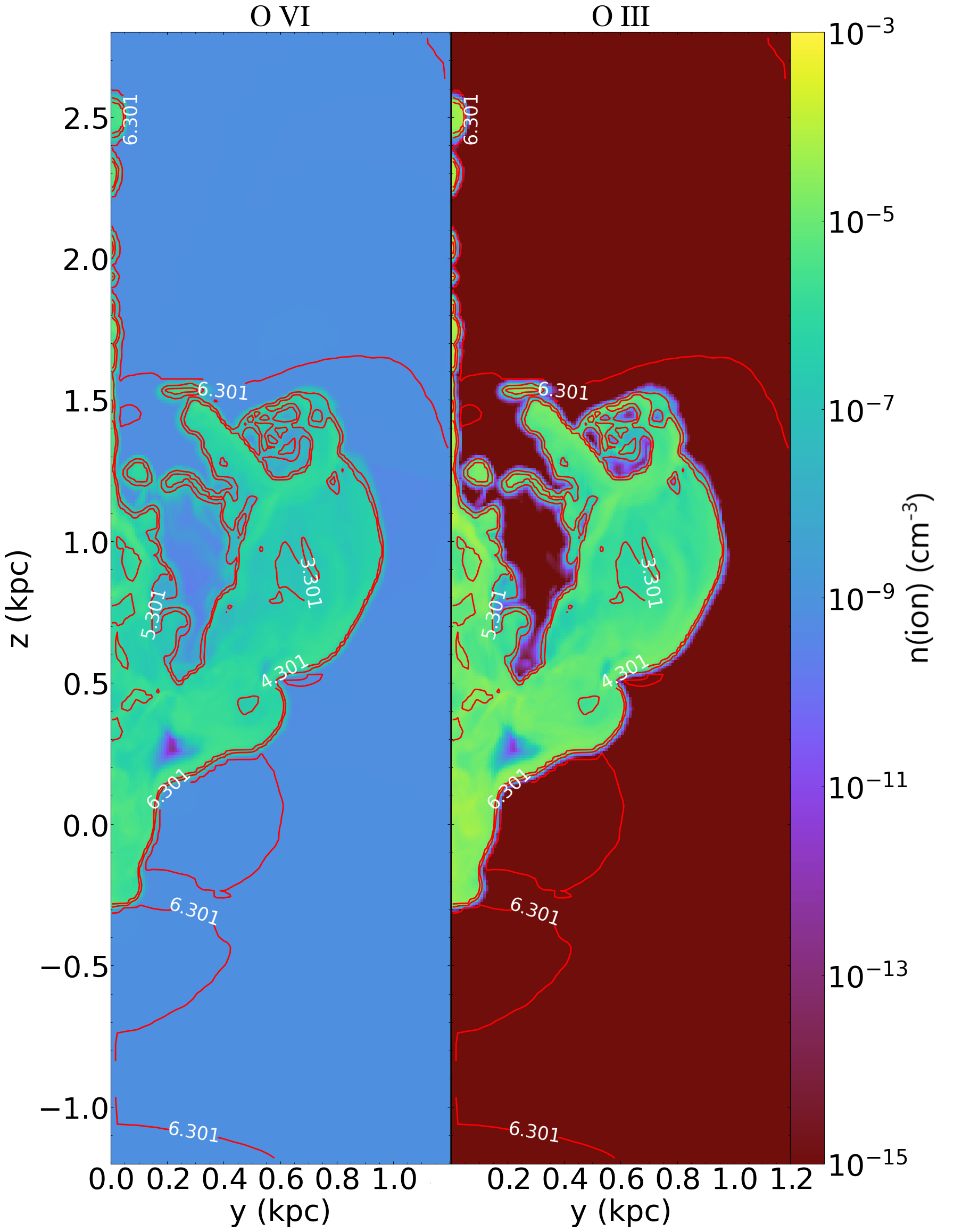}
    \caption{The number density and temperature of the gas in the domain at $x = 0$ kpc. 
    The left panel is for \ion{O}{6} and the right panel is for \ion{O}{3}.
    The color bar represents the  number density and the contours show the gas temperature. The four contour levels are \mbox{$2\times10^3$ K}, \mbox{$2\times10^4$ K}, \mbox{$2\times10^5$ K} and \mbox{$2\times10^6$ K}. Substantial fractions of \ion{O}{6} and \ion{O}{3} can be seen across a wide range of temperatures, from less than \mbox{$2\times10^3$ K} to more than \mbox{$2\times10^5$ K}.
    There is significant overlap between \ion{O}{6}-rich gas and \ion{O}{3}-rich gas.}
    \label{fig:o6slice}
\end{figure*}

As a result of this mixing, ionization, and recombination, \ion{O}{6} exists throughout the cloud and at a wide range of temperatures.  The left panel of Figure \ref{fig:o6slice} shows the \ion{O}{6} spatial distribution and the temperature distribution for Run 1 at 100 Myr.  Due to their dynamic nature, our set of simulations contains \ion{O}{6} across a wide temperature range of $\sim 2,000$ K to $\sim 2,000,000$ K. In contrast, static CIE and NEI models, such as those of \citet{Sutherland93} and \citet{GS07}, predict that \ion{O}{6} exists in a comparatively narrow temperature range of $\sim 150,000$ K to $\sim 1,000,000$ K.

\begin{figure}[h]
    \centering
    \includegraphics[scale=1.0]{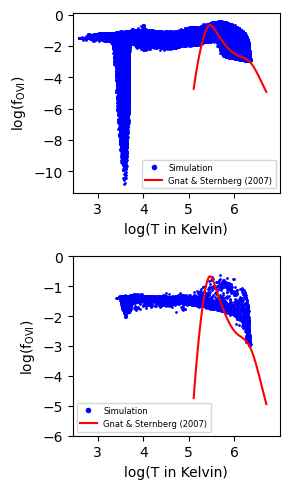}
    \caption{Top panel:   the $f_{\rm OVI}$ and temperature are plotted with a blue dot for every cell in the cloud. 
Bottom panel:   the $f_{\rm OVI}$ and temperature are plotted with a blue dot for every sight line
through the cloud.   The sight lines are oriented parallel to the $x$ axis and the
temperature is the mass-weighted average temperature along the line of sight. 
Both plots were made from Run 1 at 100 Myr.    Only the material that met the
cloud’s velocity criterion was used to make these plots.  For comparison, the $f_{\rm OVI}$ vs T curve for static gas is plotted in red.   It was adopted from \citet{GS07}.
\ion{O}{6} is confined to a narrower range of temperatures in the static curve than in
our simulations, which include dynamic mixing of warm and hot gas.}
    \label{fig:f4}
\end{figure}

\begin{figure}[ht]
    \centering
    \includegraphics[scale=1.0]{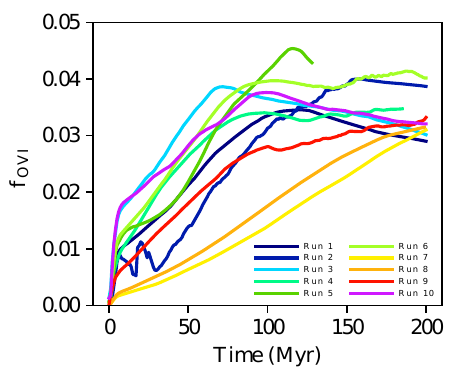}
    \caption{Plot of the cloud-averaged $f_{\rm OVI}$ for each simulation as a function of time. 
 Generally, $f_{\rm OVI}$ increases with time as the cloud mixes with the ambient material.}
    \label{fig:ifo6}
\end{figure}

The fact that we find \ion{O}{6} at a wide range of temperatures is important because it means that \ion{O}{6} does not solely trace hot, ionized hydrogen, but instead exists in all phases of gas. It also has implications for the calculation of the hydrogen column density, which will be discussed later in this section.  
Furthermore, the oxygen atoms in the \ion{O}{6} state, $f_{\rm OVI}$, is significant across the entire temperature range found in the cloud.  This is true both when examining cells and when examining sight lines through the clouds (see Figure \ref{fig:f4}).  The evolution of the $f_{\rm OVI}$ is shown in Figure \ref{fig:ifo6}.  At the beginning of each simulation, $f_{\rm OVI}$ is small. Over time, the cloud mixes with the ambient gas, causing $f_{\rm OVI}$ to rise. At even later epochs, $f_{\rm OVI}$ stabilizes. 

For comparison with observations, we estimate the column density of \ion{H}{1} using the following method. Although FLASH cannot distinguish \ion{H}{1} from \ion{H}{2} in each cell, it can track the ionization and recombination of other elements.  We therefore use \ion{O}{1} to trace \ion{H}{1}.
We calculate the quantity of \ion{H}{1} in collisionally ionized gas from the ratio of \ion{O}{1} to all oxygen in the cell times the quantity of hydrogen in each cell. This calculation is justified because the ratio of \ion{H}{1} to all hydrogen is approximately equal to the ratio of \ion{O}{1} to all oxygen, since oxygen and hydrogen have very similar first ionization potentials \citep{Allen} and charge exchange between \ion{O}{2} and \ion{H}{1} particles is common \citep{Field71}. We use this procedure to calculate the column densities of neutral hydrogen for every sightline through the HVC; these are the column densities in collisionally ionized gas. Then, in order to approximate the effect of photoionization, we subtract from these column densities the column density of hydrogen that has been photoionized.  We use the resulting \ion{H}{1} to calculate the ratio of \ion{H}{1} to \ion{O}{6}.

\begin{figure}[ht]
    \centering
    \includegraphics[scale=1.0]{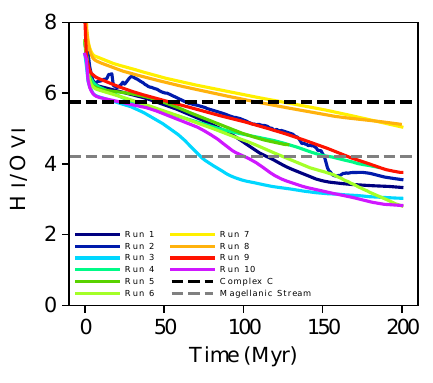}
    \caption{Plots of \ion{H}{1}/\ion{O}{6} for each simulation as a function of time.  Generally, \ion{H}{1}/\ion{O}{6} decreases with time as hydrogen is ionized and $f_{\rm OVI}$ increases (see Figure \ref{fig:ifo6}).  Also shown are the average observed {\ion{H}{1}/\ion{O}{6}} ratios for Complex C and the MS \citepalias{Sembach03,Fox10}.}
    \label{fig:h1o6}
\end{figure}

For each epoch in a simulation, we calculate the ratio of \ion{H}{1} to \ion{O}{6} over the entire cloud. These curves can be seen in Figure \ref{fig:h1o6}. This ratio is highest at the start of the simulation; this is because there has been little mixing and so the cloud consists of mostly cool, neutral gas. As the simulation evolves, the cloud mixes with the ambient gas, causing this ratio to decline until it reaches an asymptote. For comparison, the ratio of \ion{H}{1} to \ion{O}{6} in the MS and Complex C are $1.7 \times 10^4$ \citepalias{Fox10} and $5.6 \times 10^5$  \citepalias{Sembach03}. These two values are shown in Figure \ref{fig:h1o6} as horizontal lines.  Our simulations best replicate the MS and Complex C during the epochs when they have the same \ion{H}{1}/\ion{O}{6} ratios as these clouds.

\begin{table}[!ht]
    \caption{  { $f_{\rm OVI}$}}\label{tab:notes}
    \centering 
    \hspace*{-3cm}\begin{threeparttable}
    \begin{tabular}{lccc}\toprule
      Clouds & Sight lines & $f_{\rm OVI}$\tnote{a}\\
      \bottomrule
      Complex C & Mrk 279 &   { 0.023} \\ 
       & Mrk 290 &   {0.019} \\ 
       & Mrk 501 &   {0.031} \\ 
       & Mrk 506 &   {0.036} \\ 
       & Mrk 817 &   {0.023} \\ 
       & Mrk 876 &   {0.031} \\ 
       & PG 1259+593 &   {0.0099} \\ 
       & PG 1351+640 &   {0.014} \\ 
       & PG 1626+554 &   { 0.031} \\ 
       & Average $f_{\rm OVI}$\tnote{b} &   {0.024} \\ 
       & $f_{\rm OVI}$ from average \ion{H}{1}/\ion{O}{6}\tnote{c} &   {0.020} \\
      \bottomrule
      Magellanic & NGC 7469 &   {0.035} \\ 
      Stream & Mrk 335 &   {0.034} \\
      &   {HE 0226-4110} &   {0.035} \\
       &   {Average $f_{\rm OVI}$\tnote{d}} &   {0.035} \\ 
       &   {$f_{\rm OVI}$ from average \ion{H}{1}/\ion{O}{6}\tnote{e}} &   {0.036} \\ 
      \bottomrule

    \end{tabular} 
    \begin{tablenotes}
       \item[a] The $f_{\rm OVI}$ for each individual sight line is calculated using the \ion{H}{1}/\ion{O}{6} of individual sight lines from \citetalias{Sembach03} and \citetalias{Fox10}. 
         It is the average of the values obtained from the ten simulations.
       \item[b] This $f_{\rm O VI}$ is the average value of the $f_{\rm O VI}$ of the nine sight lines in Complex C.  It is calculated from the unrounded values.
       \item[c] This $f_{\rm O VI}$ is the average value of those calculated from the simulations at the epochs when   {\ion{H}{1}/\ion{O}{6}} matches the mean value of \ion{H}{1}/\ion{O}{6} for the 9 sight lines listed.     
       \item[d] This $f_{\rm O VI}$ is the average value of the $f_{\rm O VI}$ of the three sight lines in the MS.  It is calculated from the unrounded values.
       \item[e] This $f_{\rm O VI}$ is the average value of those calculated from the simulations at the epochs when (\ion{H}{1}/\ion{O}{6}) matches the mean value of \ion{H}{1}/\ion{O}{6} for the three sight lines listed.

     \end{tablenotes}
    \end{threeparttable}\hspace*{-3cm}
    \label{tab:fovi}
  \end{table}
  
Table \ref{tab:fovi} lists the 12 sight lines through Complex C and the MS for which \ion{H}{1} and \ion{O}{6} have been measured by   \citetalias{Sembach03} and \citetalias{Fox10}.  We use the ratios of these observed \ion{H}{1} and \ion{O}{6} column densities to determine the most appropriate epochs in the simulations for further analysis. 
We then determine the value of $f_{\rm OVI}$ at the corresponding epoch for each simulation for each sight line, using the values plotted in Figure \ref{fig:ifo6}. Note that $f_{\rm OVI}$ is not very sensitive to the choice of epoch.  We then determine the average value of $f_{\rm OVI}$ for each sight line by averaging the corresponding $f_{\rm OVI}$s from the ten simulations at the epochs when their \ion{H}{1}/\ion{O}{6} ratio matches that of the sight line. 
Our results are shown in Table \ref{tab:fovi}.  Our ionization fractions are significantly lower than the maximum value of 0.22 predicted from CIE and NEI models of static gas \citep{Sutherland93,GS07}. 
Until now, the maximum $f_{\rm OVI}$ in static gas has been used to determine the mass of hot ionized hydrogen in HVCs \citepalias{Fox10,Sembach03}. The ratio of the maximum $f_{\rm OVI}$ in static gas to our average $f_{\rm OVI}$ in dynamic gas ranges from \mbox{6.3 (Mrk 335)} to \mbox{22.2 (PG1259+593)}.  This result leads to a related increase in the mass of HVCs, as shown below. 

\subsection{New Methodology for Calculating N(H)}

Until now the typical method for calculating the amount of ionized hydrogen associated with observed column densities of \ion{O}{6} has been:

\begin{equation}\label{eq:old}
   N(\text{\ion{H}{2}}) = \frac{N(\text{\ion{O}{6}})}{\left(\frac{\text{O}}{\text{H}}\right)\left(\frac{\text{\scriptsize{\ion{O}{6}}}}{\text{O}}\right)}
\end{equation}

\noindent
e.g., \citepalias{Sembach03, Fox10} where N(\ion{O}{6}) is the column density of \ion{O}{6}, (O/H) is the metallicity, and (\ion{O}{6}/O) is $f_{\rm OVI}$. 
Without specific knowledge of the temperature distribution of gas in a cloud, it is common for researchers to adopt $f_{\rm OVI}$ as less than or equal to the maximum value in the theoretical $f_{\rm OVI}$  curve for static gas.
When \citetalias{Fox10} used this method for the MS, they adopted $f_{\rm OVI}$ $\leq 0.22$ from a static gas model \citep{GS07}. 
Figure \ref{fig:f4} compares $f_{\rm OVI}$ vs temperature for that static model with our dynamic results. In the static model, \ion{O}{6} only exists in a narrow temperature range around \mbox{$10^{5.5}$ K}, where nearly all of the hydrogen is expected to be ionized. 
Hence, typically the left side of Equation \ref{eq:old} is N({\ion{H}{2}}) rather than N(H) (e.g., \citetalias{Fox10}).
\citetalias{Sembach03} adopted \mbox{{$f_{\rm OVI}$} $\leq 0.2$} in their analysis in Complex C and used Equation \ref{eq:old}.

\citetalias{Fox10} also calculated the column densities of ionized hydrogen associated with \ion{Si}{4} and low ions. 
Their \ion{H}{2} column densities associated with \ion{Si}{4} and \ion{O}{6} are also lower limits because the ionization fractions used are upper limits.
Recognizing that there is a wide range of temperatures in a cloud, they then summed the \ion{H}{2} column densities associated with low ions, \ion{Si}{4}, and \ion{O}{6} to achieve the total \ion{H}{2} column density.

They made this summation because they assume that the low ions, \ion{Si}{4}, and \ion{O}{6} trace different temperature regimes and therefore different populations of \ion{H}{2}.  However, the turbulent mixing layer simulations of \citet{Kwak10} show that this is not always the case.  Individual cells can have populations in several ionization states.
Likewise, an example side-by-side comparison of \ion{O}{3} and \ion{O}{6} from our fiducial run (Figure \ref{fig:o6slice}) shows that the spatial distributions of intermediate and high ions greatly overlap.
In addition, 97\% of the cloud material has an \ion{O}{6} volume density of $10^{-9}$~cm$^{-3}$ in the same simulation and epoch.

These points raise the question of how well adding temperature phase components reproduces the total amount of hydrogen. 
As a simple example, consider a cloud in which there is an equal amount of oxygen in each ionization state. 
For the sake of argument, the total amount of hydrogen might be calculated in the following way:

\begin{equation*}
N(\text{H})_{\rm calc} = \sum_{i=1}^9 \frac{N(\text{O}_i)}{Z f_{i,max}},
\end{equation*}
  {where $N(\text{O}_i)$ is the column density of the ith oxygen ion, $Z$ is the oxygen metallicity, and {{$f_{i,max}$}} is the {maximum} ionization fraction for the ith ionization state of oxygen.  
Note that this {{${f_{i,max}}$}} is different from the actual fraction of ions in the ith ionization state in this example, which is $\frac{1}{9}$.
Since $Z$ is a constant and $N(\text{O}_i)$ is equal to $\frac{1}{9}$th of the total oxygen column density, $N(\text{O})$, the above equation reduces to} 

\begin{equation*}
    N(\text{H})_{\rm calc} = \frac{N(\text{O})}{9Z}\sum_{i=1}^9 \frac{1}{ f_{i,max}}.
\end{equation*} 
  {Since $\frac{N(\text{O})}{Z}$ is equal to the actual column density of hydrogen, $N(\text{H})_{\rm act}$, the expression further reduces to:}

\begin{equation*}
    N(\text{H})_{\rm calc} = \frac{N(\text{H})_{\rm act}}{9}\sum_{i=1}^9 \frac{1}{f_{i,max}}.
\end{equation*}
If {{$f_{i,max}$}} were set to the maximum ionization fractions for oxygen from the CIE tables of \citet{GS07}, then \mbox{$\sum_{i=1}^9 \frac{1}{f_{i,max}} = 15.3$}, so \mbox{$N(\text{H})_{\rm calc} = 1.70 N(\text{H})_{\rm act}$.}
This technique overcounts the actual hydrogen column density by $70\%$.

Of course, it is unlikely that all 9 ionization states of oxygen would be observed for a sight line. 
A more practical example would be to use just a few ions that are expected to span the entire temperature range of the cloud.  As an example, consider \ion{O}{1}, \ion{C}{2}, \ion{Si}{4}, and \ion{O}{6}.
In this example, the total calculated hydrogen column density might be given as:

\begin{multline*}
    N(\text{H})_{\rm calc} =
\frac{N(\text{\ion{O}{1}})}{Z_O f_{\rm O I, max}} +\frac{N(\text{\ion{C}{2}})
}{Z_C f_{\rm C II, max}} + \\
\frac{N(\text{\ion{Si}{4}})
}{Z_{Si} f_{\rm Si IV, max}} + \frac{N(\text{\ion{O}{6}})
}{Z_O f_{\rm O VI, max}},
\end{multline*}
\noindent where $Z_O$, $Z_C$, and $Z_{Si}$ are the oxygen, carbon, and silicon metallicities, respectively.
As has been done frequently, the ionization fractions used here would be the maxima from \citet{GS07}.
This calculation assumes that each ion is found in a disjoint temperature range and traces distinct gas.
If each temperature range is in CIE at the temperature that maximizes the ionization fraction from static CIE gas models, then this equation would accurately capture the total hydrogen column density along the sight line.
However, our dynamic simulations show that different ions can overlap significantly in temperature.
There is also observational evidence of high ions such as \ion{O}{6} existing at a wide range of temperatures (see Section \ref{sec:discussion}). 
Therefore, it is not accurate to assume that each ion traces different populations of gas.
To account for this, we can define an ionization fraction of \ion{O}{6} that considers \ion{O}{6} at all temperatures along a sight line as the total number of \ion{O}{6} ions divided by the total number of oxygen atoms ($f_{\rm O VI, act}$).
Suppose that for a given sight line, $f_{\rm O VI, act}$ differs from the maximum ionization fraction predicted from the static gas models of \citet{GS07} by a factor of $\beta_{\rm O VI}$, i.e., \mbox{$f_{\rm O VI, act} = \beta_{\rm O VI}f_{\rm O VI, max}$}, and likewise for the ionization fractions of \ion{O}{1}, \ion{C}{2} and \ion{Si}{4}.
Then, the previous equation can be re-written as:

\begin{multline*}
    N(\text{H})_{\rm calc} = 
\frac{N(\text{\ion{O}{1}})\beta_{\rm O I}
}{Z_O f_{\rm O I, act}} + \frac{N(\text{\ion{C}{2}})\beta_{\rm C II}
}{Z_C f_{\rm C II, act}} \\
+ \frac{N(\text{\ion{Si}{4}})\beta_{\rm Si IV}
}{Z_{Si} f_{\rm Si IV, act}} + \frac{N(\text{\ion{O}{6}})\beta_{\rm O VI}
}{Z_O f_{\rm O VI, act}}. 
\end{multline*}

However, $\frac{N(\text{\scriptsize{\ion{O}{6}}})}{f_{\rm O VI, act}}$ is by definition equal to the actual column density of oxygen along the entire sight line, $N(\text{O})_{\rm act}$.  Additionally, since $Z_O$ is the oxygen abundance along the sight line, $\frac{N(\text{O})_{\rm act}}{Z_O}$ is equal to the total hydrogen column density along the sight line, $N(\text{H})_{\rm act}$.  The same argument applies to \ion{O}{1}, \ion{C}{2} and \ion{Si}{4}.
This allows a further simplification:

\begin{equation*}
    N(\text{H})_{\rm calc} = N(\text{H})_{\rm act} \left(\beta_{\rm O I} + \beta_{\rm C II} + \beta_{\rm Si IV} + \beta_{\rm O VI} \right).
\end{equation*}

This example highlights the problem with using multiple ions and assuming maximum ionization fractions.
As a hypothetical example, consider a sight line that contains a substantial amount of \ion{O}{6}, such that $\beta_{\rm O VI} = \frac{1}{2}$, and lesser amounts of \ion{O}{1}, \ion{C}{2}, and \ion{Si}{4}, such that \mbox{$\beta_{\rm O I} = \beta_{\rm C II} = \beta_{\rm Si IV} = \frac{1}{4}$}.
These choices of values may be realistic for a sight line that contains significant portions of hot gas.
Using those $\beta$ values, the sum of the four actual ionization fractions would be \mbox{$\sim0.64$}, indicating that a substantial percentage of the metals along the sight line are in those ionization states.
In this case, this method would overcount the total hydrogen column density by a factor of 1.25.
On the other hand, if each ion's ionization fraction is smaller than its CIE maximum value by a factor such as \mbox{6.3--22.2}, which are the factors found from our simulations for \ion{O}{6}, then this method results in undercounting by a factor of 1.6--5.6.

The above logic would need only slight modification if Cloudy is used to find the low ion column density and the neutral hydrogen column density is observed directly rather than deduced via \ion{O}{1}, as in \citetalias{Fox10}.

From observations alone, it is not possible to obtain values for $\beta$, so it would not be clear whether the calculated hydrogen column density is over or under counting the actual hydrogen column density. 
However, dynamic NEI computer simulations make it possible to calculate the ionization fraction of \ion{O}{6} (or any other ion) in simulated clouds.  
This ionization fraction takes into account \ion{O}{6} found throughout the whole cloud and therefore coincident with neutral and ionized hydrogen.  As a result, we remove the assumption that \ion{O}{6} only traces hot, ionized gas in Equation \ref{eq:old} by replacing N(\ion{H}{2}) with N(H) and using the \ion{O}{6}/O from our simulations.
The result is Equation \ref{eq:new}, which is more mathematically accurate since the metallicity is the ratio of the total amount of oxygen to the total amount of hydrogen, not just ionized hydrogen.

\begin{equation}\label{eq:new}
   N(\text{H}) = \frac{N(\text{\ion{O}{6}})}{\left(\frac{\text{O}}{\text{H}}\right)\left(\frac{\text{\scriptsize{\ion{O}{6}}}}{\text{O}}\right)}.
\end{equation}


\begin{table*}[!ht] 
    \caption{Complex C { Log} Column Densities}\label{tab:comc}
    \centering 
    \hspace*{-3cm}\begin{threeparttable}
    \begin{tabular}{lcccccccc}\toprule
      Sight Lines & N({\ion{O}{6}})\tnote{a} & N({\ion{H}{1}})\tnote{a} & N({\ion{H}{2}})\tnote{b} & N({\ion{H}{1}})+N({\ion{H}{2}}) & N(H)\tnote{c} & N(H){$_{\rm alt}$}\tnote{d}\\ 
                 & (cm$^{-2}$) & (cm$^{-2}$) & (cm$^{-2}$) & (cm$^{-2}$) & (cm$^{-2}$)& (cm$^{-2}$)\\ 
                 
      \bottomrule
      Mrk 279 & 13.67 & 19.31 & {${>}$}18.68 & {${>}$}19.40 &   {19.62} &   {19.67}\\ 
      Mrk 290 & 14.20 & 19.98 & {${>}$}19.21 & {${>}$}20.05 &   {20.22} &   {20.21}\\ 
      Mrk 501 & 13.81 & 19.03 & {${>}$}18.82 & {${>}$}19.25 &   {19.63} &   {19.82}\\ 
      Mrk 506 & 14.05 & 18.64 & {${>}$}19.06 & {${>}$}19.20 &   {19.81} &   {20.06}\\ 
      Mrk 817 & 13.88 & 19.51 & {${>}$}18.89 & {${>}$}19.01 &   {19.83} &   {19.89}\\ 
      Mrk 876 & 14.05 & 19.30 & {${>}$}19.06 & {${>}$}19.50 &   {19.87} &   {20.06}\\ 
      PG 1259+593 & 13.72 & 19.95 & {${>}$}18.73 & {${>}$}19.98 &   {20.03} &   {19.73}\\ 
      PG 1351+640 & 13.75 & 19.78 & {${>}$}18.76 & {${>}$}19.90 &   {19.93} &   {19.76}\\ 
      PG 1626+554 & 14.22 & 19.43 & {${>}$}19.23 & {${>}$}19.64 &   {20.04} &   {20.23}\\
      \bottomrule

    \end{tabular} 
    \begin{tablenotes}
       \item[a] The values of N(\ion{O}{6}) and N(\ion{H}{1}) are from \citetalias{Sembach03}. 
       \item[b] The values of N(\ion{H}{2}) are calculated based on the N(\ion{O}{6}) and $f_{\rm OVI}$ from   \citetalias{Sembach03}, using the \citet{Asplund09} abundances.
       \item[c] For each sight line, the value of N(H) is calculated using N({\ion{O}{6}}) from \citetalias{Sembach03} and the calculated value of   {$f_{\rm OVI}$}   {for that sight line} in Table \ref{tab:fovi}.  
       \item[d] For each sight line, the {alternate} value of N(H) is calculated using N({\ion{O}{6}}) from   \citetalias{Sembach03} and the ``$f_{\rm OVI}$ from average \ion{H}{1}/\ion{O}{6}" in Table \ref{tab:fovi}.  
       
     \end{tablenotes} 
     \end{threeparttable}\hspace*{-3cm}
  \end{table*} 

  \begin{table*}[!ht] 
    \caption{Magellanic Stream Log Column Densities}\label{tab:ms}
    \centering
    \hspace*{-3cm}\begin{threeparttable} 
    \begin{tabular}{lccccccccc}\toprule
      Sight Lines& N(\ion{O}{6})\tnote{a} & N(\ion{H}{1})\tnote{a} & N(\ion{H}{2})$_{\text{\tiny{\ion{O}{6}}}}$\tnote{a} &   {N(\ion{H}{2})$_{\text{\tiny{low + \ion{Si}{4}}}}$\tnote{a}} & N({\ion{H}{1}})+N({\ion{H}{2}}) & N(H)\tnote{b} & N(H){$_{\rm alt}$}\tnote{c}\\ 
                 & (cm$^{-2}$) & (cm$^{-2}$) & (cm$^{-2}$) & (cm$^{-2}$)&(cm$^{-2}$) & (cm$^{-2}$)& (cm$^{-2}$)\\  
      \bottomrule
      NGC 7469 &   {14.09} & 18.63 & {${>}$}19.10 & 
      {${>}$}19.90 & {${>}$}19.98 &   {19.71} &   {19.70} \\ 
      Mrk 335 &   {13.84} & 16.67 & {${>}$}18.80 & 
      {${>}$}18.95 & {${>}$}19.18 &   {19.80} &   {19.78} \\
        {HE 0226-4110} &   {13.91} & $\approx17.0$ &   {${>}$}{18.9} & 
        {${>}$}{19.18} &   {${>}$}{19.37} &   {19.68} &   {19.66} \\
      \bottomrule

    \end{tabular} 
    \begin{tablenotes}
       \item[a] The values of N(\ion{O}{6}), N(\ion{H}{1}) and N(\ion{H}{2}) are from \citetalias{Fox10}. 
       \item[b] For each sight line, the value of N(H) is calculated using N(\ion{O}{6}) from \citetalias{Fox10} and $f_{\rm OVI}$ in Table \ref{tab:fovi}.  
       \item[c] For each sight line, the alternate value of N(H) is calculated using N(\ion{O}{6}) from \citetalias{Fox10} and the ``$f_{\rm OVI}$ from average \ion{H}{1}/\ion{O}{6}" in Table \ref{tab:fovi}.
     \end{tablenotes}
    \end{threeparttable}\hspace*{-3cm}
  \end{table*}
Since a single sight line may happen to pass through particularly cool or hot gas, we recommend using multiple sight lines when comparing with our method. 

\subsection{Comparisons}
\label{sec:comps}
To see the effect of shifting to this new methodology, we calculate N(H) for the sight lines in \citetalias{Fox10} and \citetalias{Sembach03}, using Equation \ref{eq:new}, our $f_{\rm OVI}$, and N(\ion{O}{6}) taken from the observations reported in \citetalias{Fox10} and \citetalias{Sembach03}. These values are shown in column 6 in Table \ref{tab:comc} and column 7 in Table \ref{tab:ms}. We think this is the most accurate method for calculating the total hydrogen densities, but we also provide the values calculated from the averages of the $f_{\rm OVI}$s extracted from the simulations at the epochs when the values of \ion{H}{1}/\ion{O}{6} match the average values of \ion{H}{1}/\ion{O}{6} from \citetalias{Fox10} and \citetalias{Sembach03}.

First, we compare with published estimates for Complex C.
We tabulate the observed N(\ion{H}{1}), the N(\ion{H}{2}) calculated using the methodologies of \citetalias{Sembach03}, and the sum of these column densities.  
The tabulated values of N(\ion{H}{2}) are calculated using Equation \ref{eq:old}, the upper limits on $f_{\rm OVI}$ from \citet{Sutherland93}   {and} \citet{GS07}, and the observed N(  {\ion{O}{6}}), which is also tabulated in Table \ref{tab:comc}.
\citetalias{Sembach03} calculated the total \ion{H}{2} mass of Complex C using the abundances of \citet{Holweger01}, but did not calculate individual \ion{H}{2} column densities for their sight lines.  To be consistent with our other calculations, we use the \citet{Asplund09} abundances to calculate those column densities.

For every sight line, N(H) calculated from Equation \ref{eq:new} and our $f_{\rm OVI}$ is larger than the observed N({\ion{H}{1}}) plus the N(\ion{H}{2}) calculated from Equation \ref{eq:old} and the value of $f_{\rm OVI}$ chosen by \citetalias{Sembach03}. 
On average, these column densities (column 6) are 2.6 times greater than the column density found from the observed N(\ion{H}{1}) plus the N(\ion{H}{2}) found from Equation \ref{eq:old} (column 5).
It should be noted that technically column 5 is a lower limit.

Notice that the last two sight lines for Complex C have very similar calculated N(H). However, \mbox{PG 1626+554} is \ion{O}{6}-rich whereas \mbox{PG 1351+640} is \ion{O}{6}-poor.
Because of the low \ion{H}{1}/\ion{O}{6} ratio of \mbox{PG 1626+554}, it has a higher fraction of ionized gas than \mbox{PG 1351+640}.  
This is consistent with the \mbox{PG 1626+554} having a higher ratio of total hydrogen to neutral hydrogen than \mbox{PG 1351+640}.  
The regions sampled by these two sight lines can be thought of as in different stages of evolution, which our prescription takes into account when calculating the total amount of material.


Next, we compare with the published results for the MS.  This is more complicated because \citetalias{Fox10} also include their estimated \ion{H}{2} column density associated with low ions and \ion{Si}{4}.  These numbers are tabulated in Table \ref{tab:ms}.  
For the \ion{H}{2} associated with \ion{Si}{4}, \citetalias{Fox10} did a similar calculation as for \ion{O}{6}, but using the \ion{Si}{4} column densities, metallicity, and maximum CIE ionization fraction.  
For the \ion{H}{2} associated with low ions,   \citetalias{Fox10} used Cloudy to model the degree of ionization in the cloud, assuming that the low ions come from photoionization rather than NEI collisional ionization. 
The \ion{H}{2} column density associated with low ions comes from the best-fitting model.  
The total hydrogen column density predicted by \citetalias{Fox10} is the sum of \ion{H}{1} and the \ion{H}{2} associated with low ions, \ion{Si}{4}, and \ion{O}{6}. 

For the three sight lines listed in Table \ref{tab:ms}, our N(H) values (column 7) are an average of 2.2 times larger than the totals in column 6 found from \citetalias{Fox10}. 
As with column 5 in Table \ref{tab:comc}, column 6 in Table \ref{tab:ms} contains lower limits.
For the first sight line, NGC 7469, our result is actually smaller than that found from \citetalias{Fox10}.  
That sight line is richer in low ions and \ion{Si}{4} than high ions and therefore has a low predicted \ion{O}{6} column density. 
Since our method is solely based on the observed \ion{O}{6} column density, it results in a low hydrogen column density.
The other two sight lines are tilted more towards \ion{O}{6}, and so our method produces higher hydrogen column densities.

\begin{figure}[ht]
    \centering
    \includegraphics[scale=1.0]{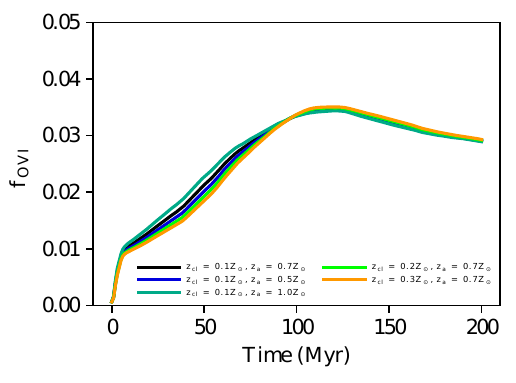}
    \caption{Plot of $f_{\rm OVI}$ for Run 1 for a variety of cloud and ambient metallicities.  Changing the ambient metallicity has little effect on the curve.  Changing the cloud metallicity has a more noticeable but still insignificant effect.}
    \label{fig:ifo6cm}
\end{figure}

\begin{figure}[ht]
    \centering
    \includegraphics[scale=1.0]{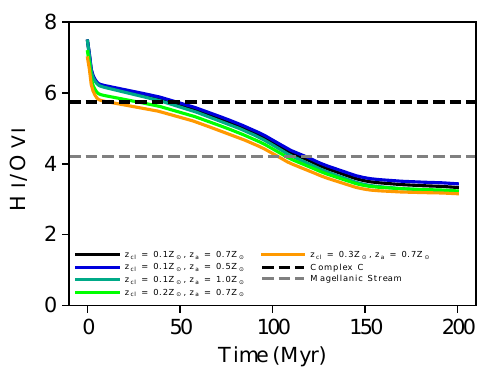}
    \caption{Plots of \ion{H}{1}/\ion{O}{6} for Run 1 for a variety of cloud and ambient metallicities.  As for $f_{\rm O VI}$, changing the ambient metallicity has little effect on the curve.  Changing the cloud metallicity has a more noticeable, but still small effect.}
    \label{fig:h1o6cm}
\end{figure}

  {We now examine the effect of metallicity.
Changing the metallicity of the cloud or ambient material shifts the \ion{H}{1}/\ion{O}{6} vs time curve, changing the epoch that best matches observations. This changes the corresponding $f_{\rm OVI}$ value that is used to calculate the hydrogen column density.  
In the case of increasing the cloud's metallicity, the effect is to decrease the appropriate $f_{\rm OVI}$.  However, since $f_{\rm OVI}$ is multiplied by the cloud metallicity when calculating the hydrogen column density, the decrease in $f_{\rm OVI}$ is mitigated by the increase in metallicity.  The net result is a small change to the hydrogen column density.
In the case of changing the ambient metallicity, the effect on the \ion{H}{1}/\ion{O}{6} vs time curve is tiny, resulting in no significant change to the choice of $f_{\rm OVI}$ or calculated hydrogen column density.  See Figures \ref{fig:ifo6cm} and \ref{fig:h1o6cm}.}

\section{Discussion} \label{sec:discussion}
We find that the \ion{O}{6}/oxygen ratio is greatly affected by dynamic mixing between hot and cool gas.  NEI ionization and recombination are much more important in mixed gas than in static gas, resulting in a significantly different $f_{\rm OVI}$.
The $f_{\rm OVI}$ for static gas is substantial only for a narrow range of temperatures and peaks at a value of 0.22 at a temperature of $3\times10^5$ K.  
This value is often used as an upper limit in calculations. In contrast, the   {$f_{\rm OVI}$} in our simulations is substantial across 3 orders of magnitude in temperature, from \mbox{$\sim2\times10^3$ K} to \mbox{$\sim2\times10^6$ K}.
Considering that the actual value of $f_{\rm OVI}$ varies over time and space, we adopt the value of $f_{\rm OVI}$ averaged over the whole simulated cloud at the time when the simulated \ion{H}{1}/\ion{O}{6} ratio best matches the ratio for the observed sight line. 

For any given sight line, we have 10 values of $f_{\rm OVI}$, one from each simulation.  For each sight line, we compare the $f_{\rm OVI}$ values from the 10 simulations.  
Amongst the simulations, the standard deviation of $f_{\rm OVI}$ is small -- roughly 5-13\% of the average value across the 10 simulations.  This suggests that the choice of initial parameters for the simulations does not have a large effect on the final results. 
Our average $f_{\rm OVI}$s are 4.5\% - 15.9\% of the peak $f_{\rm OVI}$ for static gas.


Historically, $f_{\rm OVI}$ has been used to estimate the quantity of hydrogen associated with the observed \ion{O}{6}.  Our simulations show that \ion{O}{6} can be found in cool gas, where neutral hydrogen can exist, so we revise the methodology such that it predicts the sum of the neutral and ionized hydrogen.  
The differences between our method (Equation \ref{eq:new}) and the previous method (Equation \ref{eq:old}) are that 1.) our result is the total hydrogen column density including the neutral hydrogen, while the previous method attributes the resulting column density to only the ionized hydrogen around a temperature of $3\times10^5$ K; 2.) we use the average of $f_{\rm OVI}$ of the simulated cloud instead of the peak value of $f_{\rm OVI}$ in CIE gas; and 3.) our method is an equality while the previous method provides a lower limit. 

We use our method to estimate N(H) for sight lines through Complex C and the MS from the observed N(\ion{O}{6}) \citepalias{Sembach03, Fox10}. Our values of N(H) are 2.6 and 2.2 times greater than the sum of the observed N(\ion{H}{1}) and the N(\ion{H}{2}) calculated using the older methodologies described in Section \ref{sec:comps}. 
It should be noted that the quoted N(\ion{H}{2}) from the MS is the sum of multiple components.

Stationary CIE and NEI gas has a very different $f_{\rm OVI}$ than dynamic NEI gas.  This difference has several implications.
First, as shown above, there should be more material in HVCs than previously thought.  Our column densities are noticeably higher than previous lower limits; therefore, our cloud masses would be correspondingly higher as well.  This logic should apply to any HVC for which the \ion{H}{2} column density is calculated from the \ion{O}{6} column density. 
 
Another implication pertains to {low-redshift} extragalactic \ion{O}{6} absorbers.
\citet{Sembach04} analyzed 25 {low-redshift extragalactic} \ion{O}{6} absorbers along the line of sight to PG 1116+215, finding that their ionization pattern was similar to that of HVCs.  
Furthermore, our finding that a significant portion of the O~VI is in gas with temperatures below \mbox{$10^5$ K} (see Figure \ref{fig:f4}) also agrees with observations of {low-redshift} extragalactic \ion{O}{6} systems.
\citet{Tripp08} analyzed 51 \ion{O}{6} systems, finding that at least a third of the intervening \ion{O}{6} components ``present compelling evidence of cool temperatures \mbox{($\log(T) < 5.0$)}."  
This suggests that these systems are out of equilibrium.
Another example of cool \ion{O}{6} is from \citet{Savage14}, who examined 14 sight lines with Hubble Space Telescope's Cosmic Origins Spectrograph (HST/COS). They saw 54 {low-redshift} \ion{O}{6} absorber systems, which they decomposed into 85 components. They discussed the \ion{O}{6} absorber components that aligned well with the \ion{H}{1} components. Among the 45 well-aligned components, 31 have narrow line widths, implying values of \mbox{log(T) $<$ 4.8}. This is much smaller than the CIE temperature, \mbox{log(T) = 5.5}. Traditionally, such narrow lines were thought to be due to the photoionization, but mixing can result in regions where the gas temperature is much less than the CIE temperatures of ions present in the gas \citep{Kwak10}. These examples show that significant amounts of \ion{O}{6} in {low-redshift} intergalactic absorbers are out of CIE. Therefore we use our NEI $f_{\rm OVI}$ to revise the estimate of the baryonic content of these systems.

 Based on our $f_{\rm OVI}$, there should be more material in {low-redshift} extragalactic regions than previously thought.  
 The amount of {low-redshift} extragalactic material in these \ion{O}{6} systems has been estimated from observations of the \ion{O}{6} column density, like was done with HVCs \citep{TrippSavage00, Tripp00, Sembach04}.
In these cases, $f_{\rm OVI}$ appears in the denominator of the equation.  
Here we apply our $f_{\rm OVI}$ to those observations. {These low-redshift} extragalactic \ion{O}{6} absorbers have smaller \mbox{\ion{H}{1}/\ion{O}{6}} ratios than either the MS or Complex C.  Our simulations do not produce such low \ion{H}{1}/\ion{O}{6} ratios within the simulated timeframe, but at late times our simulations asymptote to their lowest values of \ion{H}{1}/\ion{O}{6} and their highest values of $f_{\rm OVI}$.  We therefore adopt the value of $f_{\rm OVI}$ at the last epoch from each simulation.  We then average across the 10 simulations to produce a single estimated $f_{\rm OVI}$ of 0.034.  Using this value of $f_{\rm OVI}$ in Equation 7 in \citet{Sembach04}, rather than their chosen value ($f_{\rm OVI}$ $\leq 0.2$), increases the baryonic content ($\Omega_b$) of \ion{O}{6} absorbers by up to a factor of {5.9}, to \mbox{$\Omega_b$(\ion{O}{6}) = 0.013 $h_{75}^{-1}$}.  This result implies that the amount of material in {low-redshift extragalactic} \ion{O}{6} absorption line systems is enormous.  It is several times larger than the baryonic content of the stars and gas in galaxies, i.e., $0.0032 h_{75}^{-1}$ \citep{Fukugita98,Fukugita03}.  It is also over a third of the expected baryonic content of the Universe \citep{Richter06}.   Similarly, \citet{Shull12} used hydrodynamic simulations to calculate NEI $f_{\rm OVI}$ values that are lower than those from static gas models. They then recalculated the baryonic content of the warm and hot intergalactic medium using their $f_{\rm OVI}$ and found a greater $\Omega_{b}$ than the previous value.  


We believe that our methodology provides a more accurate estimate of the hydrogen column density along sight lines through HVCs and this method can be applied to additional clouds for which \ion{H}{1} and \ion{O}{6} have been observed.   Our prescription is to use the observed \ion{H}{1}/\ion{O}{6} ratio to determine $f_{\rm OVI}$, using the equation discussed below, and then use it and the observed \ion{O}{6} column density with Equation \ref{eq:new} to determine the total hydrogen column density.  

\begin{figure}[ht]
    \centering
    \includegraphics[scale=1.0]{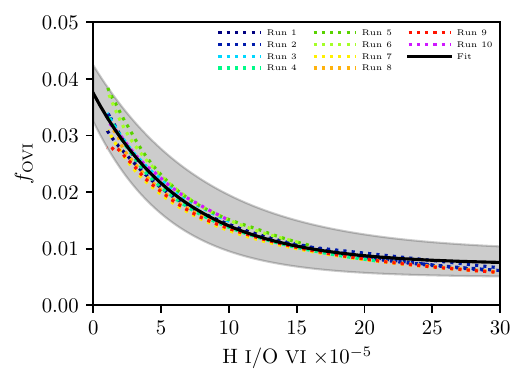}
    \caption{$f_{\rm OVI}$ vs \ion{H}{1}/\ion{O}{6} ratio for the cloud material at each epoch of the simulations.  Different colors represent different simulations.   A non-linear fit of the form \mbox{$Ae^{-bx} + C$} was performed for each simulation, where \mbox{$x$ = \ion{H}{1}/\ion{O}{6} $\times 10^{-5}$}, and the best fit values for A, b, and C were averaged.  The curve with those values  \mbox{(i.e.,   {$0.03e^{-0.17x} + 0.007$})} is plotted with a black solid line.  The shaded region depicts the 1 sigma uncertainty on the fit parameters.  This empirical relationship between $f_{\rm OVI}$ and \ion{H}{1}/\ion{O}{6} can be used to estimate a reasonable value of $f_{\rm OVI}$ from an observed \ion{H}{1}/\ion{O}{6} ratio.  Our plot extends past the typical range of observed \ion{H}{1}/\ion{O}{6} values, into a region where the fit slightly diverges from the simulation results. }
    \label{fig:pres}
\end{figure}

The simulations show that the \ion{H}{1}/\ion{O}{6} ratio is well correlated with $f_{\rm OVI}$ (see Figure \ref{fig:pres}).  The value of $f_{\rm OVI}$ falls monotonically with increasing \ion{H}{1}/\ion{O}{6}, allowing for a one-to-one relationship between observed \ion{H}{1}/\ion{O}{6} and predicted $f_{\rm OVI}$.  
Thus, an accurate $f_{\rm OVI}$ for a given sight line can be uniquely determined from the ratio of observed \ion{H}{1} and   \ion{O}{6} column density.  The relationship between $f_{\rm OVI}$ and \ion{H}{1}/\ion{O}{6} is largely independent of the input parameters of the simulations.  The best fit to these simulational data is:         
\begin{equation}
    f_{\rm OVI} = (0.030 \pm 0.003)e^{(-0.17 \pm 0.03) x} + (0.007\pm 0.002)
\end{equation}
\noindent where $x = $\ion{H}{1}/\ion{O}{6} $\times 10^{-5}$.

This analysis has been performed with oxygen. However, there are also large data sets for other elements, such as C, Si and Ne \citep{Burchett19}.  We plan to add such elements into our simulations and incorporate them in our methodology, so that hydrogen column densities can be calculated.

\section{Acknowledgements}     
We express our sincere appreciation to the anonymous referee for highlighting several important issues, leading to substantial improvements to the manuscript.  We thank M. Elliott Williams for teaching EG and CW how to run the FLASH simulations and Ashton Rutkowski for helping EG and CW with coding at the start of the project.  We acknowledge fruitful discussions concerning the calculations with Sydney Whilden.  We are grateful to Andrew Fox for providing the Milky Way radiation field for our Cloudy simulations.  We thank Shan-ho Tsai for technical assistance with the computer clusters at the GACRC.  The computer cluster used for these simulations was supported in part by resources and technical expertise from the Georgia Advanced Computing Resource Center, a partnership between the University of Georgia’s Office of the Vice President for Research and Office of the Vice President for Information Technology.  The FLASH software used in this work was developed in part by the DOE NNSA and DOE Office of Science supported Flash Center for Computational Science at the University of Chicago and the University of Rochester.  We acknowledge financial support for preliminary presentations from the University of Georgia graduate school.    

\bibliography{hvc}{}
\bibliographystyle{aasjournal}



\end{document}